\documentclass[3p,times,procedia]{elsarticle}
\usepackage{nupha_ecrc}
\usepackage{wrapfig}
\usepackage{placeins}
\usepackage{multicol}
\usepackage{amsmath}
\usepackage{mathtools}
\renewcommand{\Re}{{\rm Re}}
\renewcommand{\Im}{{\rm Im}}

\volume{00}

\firstpage{1}

\journalname{Nuclear Physics A}

\runauth{}

\jid{nupha}

\jnltitlelogo{Nuclear Physics A}

\usepackage{amssymb}
\usepackage[figuresright]{rotating}

\begin{document}
\setlength{\columnsep}{9pt}%

\begin{frontmatter}



%% Instructions from Editor: Please use the following \dochead only in the preprint version (e-print arXiv etc.); 
%\dochead{XXVIIth International Conference on Ultrarelativistic Nucleus-Nucleus Collisions\\ (Quark Matter 2018)}
\dochead{}

\title{Realistic in-medium heavy-quark potential from high statistics lattice QCD simulations}


\author[a]{P.~Petreczky}
\author[b]{A.~Rothkopf\corref{cor1}$^{*,}$ }
\author[c]{J.~Weber}

\address[a]{Physics Department, Brookhaven National Laboratory, Upton, NY 11973, USA}
\address[b]{Faculty of Science and Technology, University of Stavanger, 4021 Stavanger, Norway}
\address[c]{Department for Computational Mathematics, Science and Engineering, 
Michigan State University, East Lansing, MI, 48824, USA}

\cortext[cor1]{Speaker}

\begin{abstract}
We present our first results on a direct computation of the complex in-medium heavy quark potential from realistic lattice QCD simulations. Ensembles with $N_\tau=12$ from the HotQCD and TUMQCD collaboration offer unprecedented high statistics, those with $N_\tau=16$ unprecedented time resolution, making possible a robust extraction of the real part from the spectral functions of Wilson line correlators. To this end we deploy a combination of a Bayesian reconstruction (BR method), as well as a Pad\'e-like approximation. We corroborate findings made on less realistic lattices that $\Re[V]$ smoothly transitions from a confining to a screened behavior at high temperatures and its values lie close to the color singlet free energies. A finite value of the $\Im[V]$ is observed in the quark-gluon-plasma phase.
\end{abstract}


\end{frontmatter}

\vspace{0.5cm}

The description of heavy quarkonium properties based on a non-relativistic potential has a long history. Intuitively it is rooted in the fact that the heavy quark rest mass ($m_c=1.3$GeV, $m_b=4.7$GeV) is much larger than any other relevant scale, be it that of quantum fluctuations in QCD $\Lambda_{\rm QCD}$ or in the context of a heavy-ion collision also the scale of thermal fluctuations $T$. In turn one expects that pair production effects are highly suppressed and a non-relativistic description of the two-body system is applicable. 

In vacuum the well known Cornell model potential $V_{\rm Cornell}(r)=-\alpha(r)/r+\sigma r+c$ has been used since the late 1970's to explore the multitude of bound states below the open-heavy-flavor threshold. It incorporates the two hallmarks of QCD: asymptotic freedom from a running coupling at small distances via a Coulombic contribution, as well as confinement in the form of a linear rise at large distances. Considering the propagation of static quarks in vacuum, it is possible to link a purely real-valued potential $V_{\rm QCD}^{T=0}(r)$ between them to the exponential decay of the rectangular Wilson loop $W_{\square}(r,\tau)$ at late Euclidean times. It is straight forward to compute this quantity in lattice QCD and it turns out that $V_{\rm QCD}^{T=0}(r)$ indeed follows closely the Cornell form. Spin and velocity dependent contributions to the vacuum potential have also been determined \cite{Brambilla:2010cs}.

At finite temperature for a long time only model potentials had been available. Since at $T>0$ Euclidean time becomes compact, the relation between the late Euclidean time Wilson correlator decay and $V_{\rm QCD}^{T>0}(r)$ breaks down. Early works instead proposed to use the correlator evaluated at the latest Euclidean time $\tau=1/T$ to define a potential. This amounts to an ad-hoc identification of the color singlet free energies $F^{(1)}(r)$ with the potential. Soon after, different proposals arose based on the internal energies and linear combinations thereof. Since no Schr\"odinger equation had been derived for any of these model potentials from QCD, a long standing discussion ensued on the appropriate choice of model.

In the last decade it has become possible to give meaning to the concept of a heavy quark potential in QCD, based on the framework of effective field theory (EFT) \cite{Brambilla:2004jw}, liberating us from the need for model potentials. EFT's provide the means to systematically exploit the above mentioned separation of scales and to reformulate the dynamics of heavy quarkonium in a language of non-relativistic fields. There are two ingredients to this process, selection of the {\it relevant degrees of freedom} and {\it matching} to QCD. 

In the perturbatively constructed effective field theory potential Non-Relativistic-QCD (pNRQCD) the degrees of freedom are color singlet and color octet wavefunctions. They describe the binding properties of the two-body system, i.e. processes at the energy scale of the binding energy. Writing down the most general Lagrangian compatible with the symmetries of underlying QCD, one finds that it can be formulated in terms of time independent quantities $V^{(i)}$, ordered according to powers of the heavy quark velocity $v$, as well as gauge field dependent terms connecting singlets and octets \cite{Brambilla:2004jw}. $V^{(i)}$ are the so called Wilson coefficients of the EFT. We refer to $V^{(0)} ({\cal O}(v^0))$ as the static potential, those suppressed with $v$ as corrections.  The $V^{(i)}$ need to be determined for each realization of the hierarchy of scales. This is achieved via matching: a correlation function in the EFT is set equal to a correlation function in QCD containing the same physics content.

The situation in a non-perturbative setting is more involved, as the relation between the potential, defined as Wilson coefficient and the behavior of correlation functions entering the matching is less straight forward \cite{Burnier:2012az}. For infinitely heavy quarks we consider the following definition \vspace{-0.2cm}
\begin{align}
V_{\rm QCD}(r)=\lim_{t\to\infty}\frac{i\partial W_\square(r,t)}{W_\square(r,t)}=\lim_{t\to\infty} \frac{\int d\omega \;{\rm exp}[-i\omega t]\;\omega\; \rho_\square(r,\omega)}{\int d\omega \;{\rm exp}[-i\omega t]\; \rho_\square(r,\omega)}\label{Eq:DefPot},
\end{align}
which relates the potential among singlet states to the late Minkowski-time behavior of the Wilson loop. To be more precise, the Wilson loop is governed by the so called static energy $E^{(0)}$, which in the language of perturbative pNRQCD can only be equated with the static potential $V^{(0)}$ in the presence of a particular scale hierarchy, i.e. if $E_{\rm bind}\ll\Lambda_{\rm QCD},T$. Evaluating Eq.\eqref{Eq:DefPot} in resummed perturbation theory \cite{Laine:2006ns, Brambilla:2008cx} provided the vital insight that at $T>0$ the potential may actually contain an imaginary part. 

Note that $V_{\rm QCD}^{T>0}(r)$ does not govern the time evolution of the wavefunction but of its unequal time correlation function. An imaginary part hence is not directly related to the decay of the quarkonium state but instead to the loss of correlation between a state at initial and later time, reflecting wavefunction decoherence due to interactions with the environment \cite{Kajimoto:2017rel}.  In practice, due to the cusp divergences of the Wilson loop, we instead use Wilson line correlators in Coulomb gauge. To lowest order in HTL it has been shown that both encode the same potential and we have checked that the extraction of $\Re[V]$ is truly gauge independent.

\begin{wrapfigure}[19]{r}{0.32\textwidth}\vspace{-.9cm}
  \begin{center}
     \includegraphics[scale=0.35]{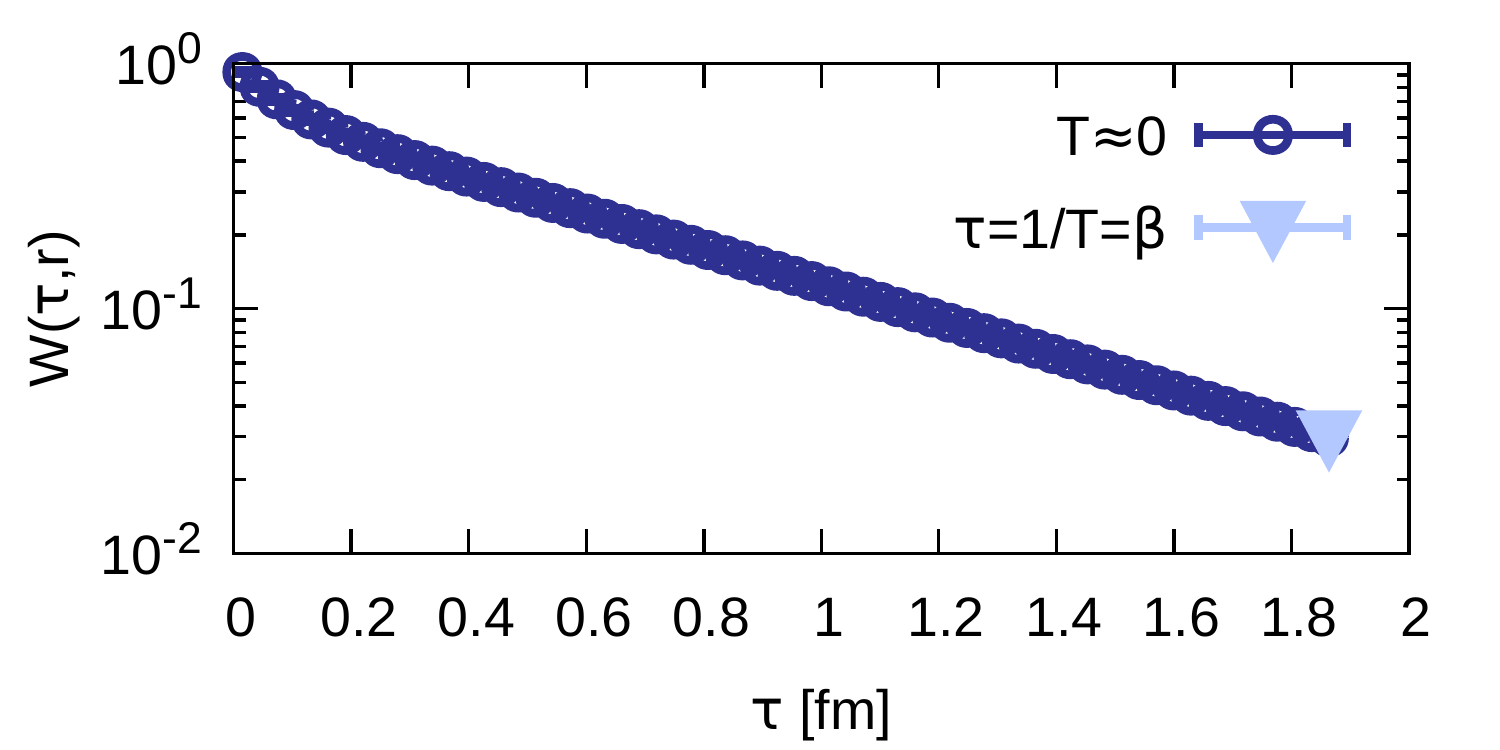}
     \includegraphics[scale=0.35]{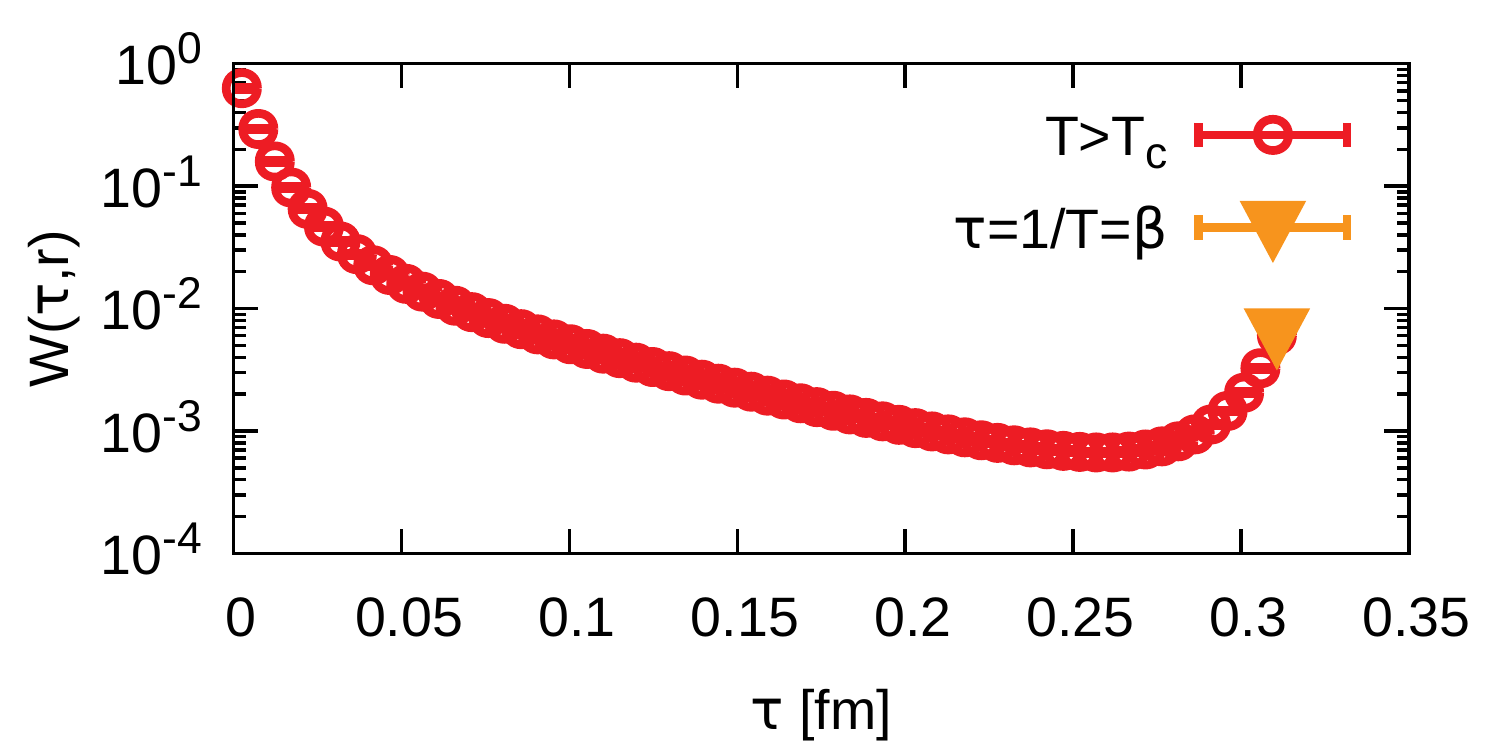}
  \end{center}\vspace{-0.7cm}
 \caption{Wilson loop at $T=0$ and $T>T_c$.}\label{Fig:DiffToOld}
\end{wrapfigure}In the temperature regime relevant for heavy-ion collisions, the potential needs to be evaluated non-perturbatively, but lattice simulations do not have direct access to the Minkowski time correlator. Instead we can use spectral functions $\rho(r,\omega)$ to bridge the Euclidean simulation and the real-time definition \cite{Rothkopf:2011db}. The same $\rho_\square(r,\omega)$ governs the Minkowski and Euclidean Wilson correlator, the former is expressed as its Fourier transform, the latter as its Laplace transform. If we have access to $\rho_\square$ we may relate it to the potential directly. Eq. \eqref{Eq:DefPot} actually tells us that if there exists a well defined lowest lying spectral peak, its position encodes $\Re[V]$ and its width $\Im[V]$. In Fig.\ref{Fig:DiffToOld} we can also see the main difference between the old potential model and the proper potential. \begin{wrapfigure}[20]{r}{0.4\textwidth}\vspace{-0.6cm}
  \begin{center}
     \includegraphics[scale=0.4, trim=0 1.1cm 0 0, clip=true ]{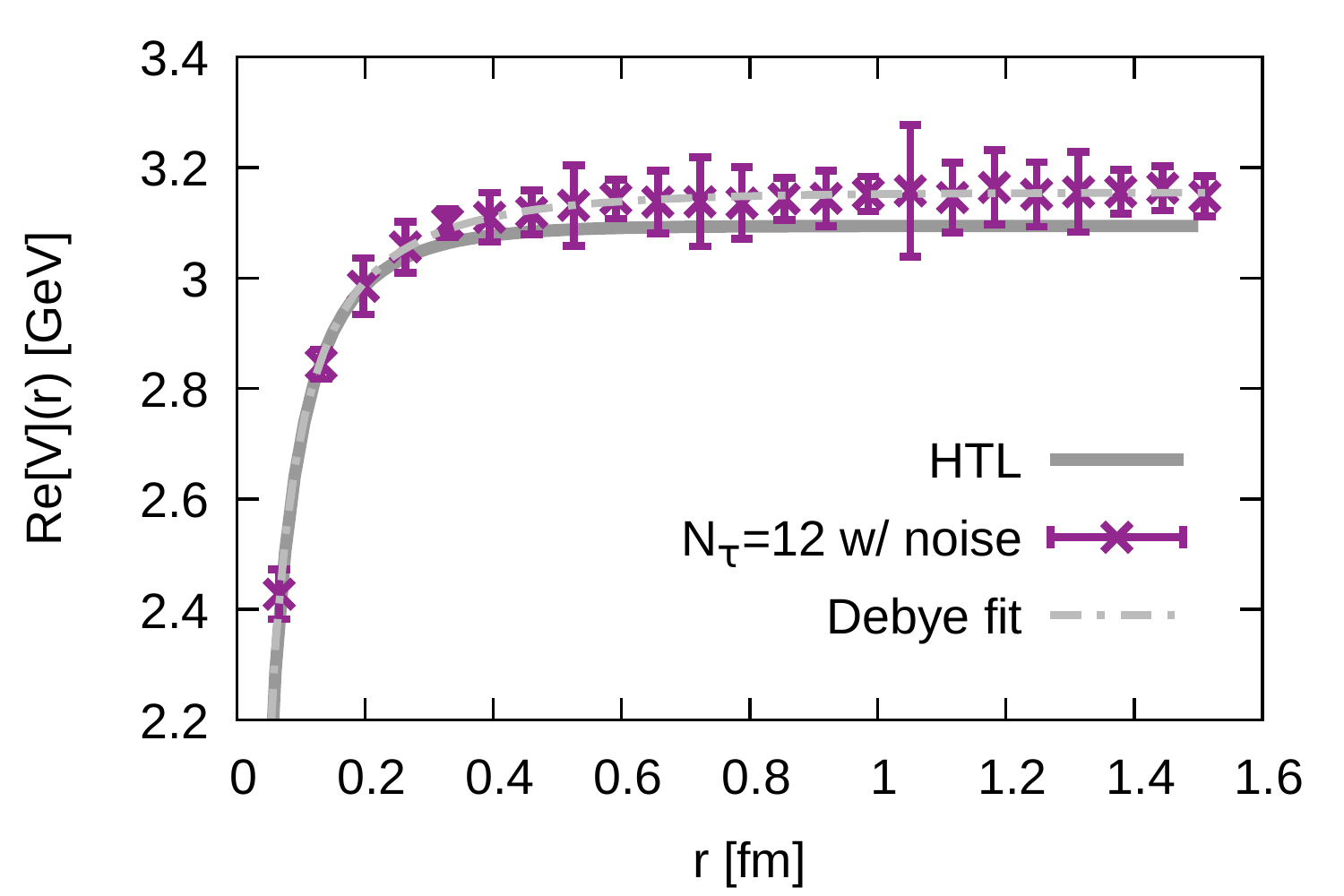}
     \includegraphics[scale=0.4]{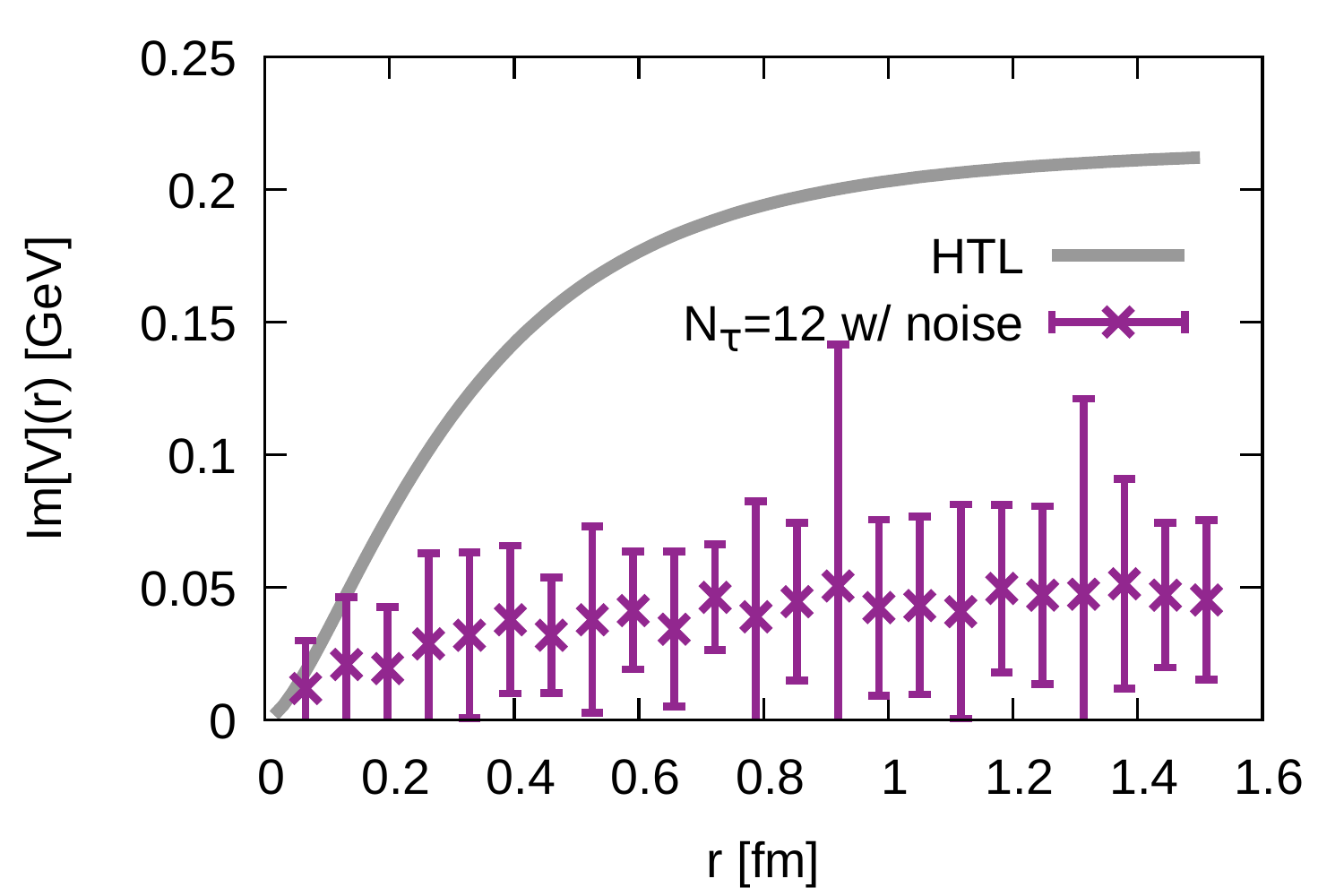}
  \end{center}\vspace{-0.7cm}
 \caption{Mock test: Pad\'e reconstructed perturbative potential using $N_\tau=12$ datapoints with $\Delta W/W=10^{-2}$.}\label{Fig:MockPade}
\end{wrapfigure}
The former only takes into account the static information encoded at $\tau=1/T$, while the latter utilizes the full information on its Euclidean time evolution. Only at $T=0$ the $\tau=1/T$ datapoint follows the same trend as the exponential falloff but at $T>0$ no obvious connection between the two remains. 

In principle $\rho_\square(r,\omega)$ can be obtained from the lattice by inverse Laplace transform. Computing it from finite and noi- sy data however is an ill-posed problem, often tackled by Bayesian inference. Here we deploy two different approaches: on the one hand the genuine Bayesian BR method \cite{Burnier:2013nla}, which exploits {\it prior information} to regularize the inversion, imposing positivity and smoothness onto the spectrum. On the other hand we deploy a rational interpolation akin to the Pad\'e approximation \cite{Tripolt:2018xeo}, which exploits the {\it analyticity} of the Wilson line correlator. One projects the simulation data in imaginary frequencies onto a set of rational functions and analytically continues these basis functions to Minkowski frequencies. Taking the imaginary part of the outcome yields $\rho_\square(r,\omega)$. We have checked the feasibility of the method using mock data, i.e. reconstructing via Pad\'e the known perturbative Wilson line spectrum and the corresponding potential. As shown in Fig.\ref{Fig:MockPade}, using a realistic $N_\tau=12$, distorted by noise with $\Delta W/W=10^{-2}$ allows us to reproduce $\Re[V]$ within uncertainties, while $\Im[V]$ is still underestimated. For $N_\tau=48$ $\Im[V]$ can be robustly extracted up to $r\approx 0.4$fm. (Note that we do not claim that the rational interpolation reconstructs the full spectrum correctly but it here allows us to recover the position of the lowest lying peak reliably.)

Here we present first results from extracting the potential from state-of-the art lattice QCD simulations by the HotQCD and TUMQCD collaboration with dynamical u,d, and s quarks, spanning a temperature range of $T\in[150,1248]$MeV. These simulations with high statistics, originally designed for the study of the QCD equation of state \cite{Bazavov:2017dsy,Bazavov:2014pvz} and screening properties \cite{Bazavov:2018wmo}, use inverse couplings $\beta=6.740\ldots9.49$. They feature an almost physical $m_\pi=161$MeV, except at $\beta=8.0,8.4$, where it is set to $m_\pi\approx 320$MeV. Ensembles with $N_\tau=12$ reach up to $T=252$MeV and $N_\tau=16$ is used beyond. For calibration, $T=0$ lattices are available up to $\beta=8.4$. The $N_\tau=12$ ensembles feature unprecedented statistics, providing 2000-9000 realizations of the Wilson correlators. For the first time $N_\tau=16$ lattice extent is available for potential extraction.
\begin{wrapfigure}[11]{l}{0.4\textwidth}\vspace{-0.85cm}
  \begin{center}
     \includegraphics[scale=0.4]{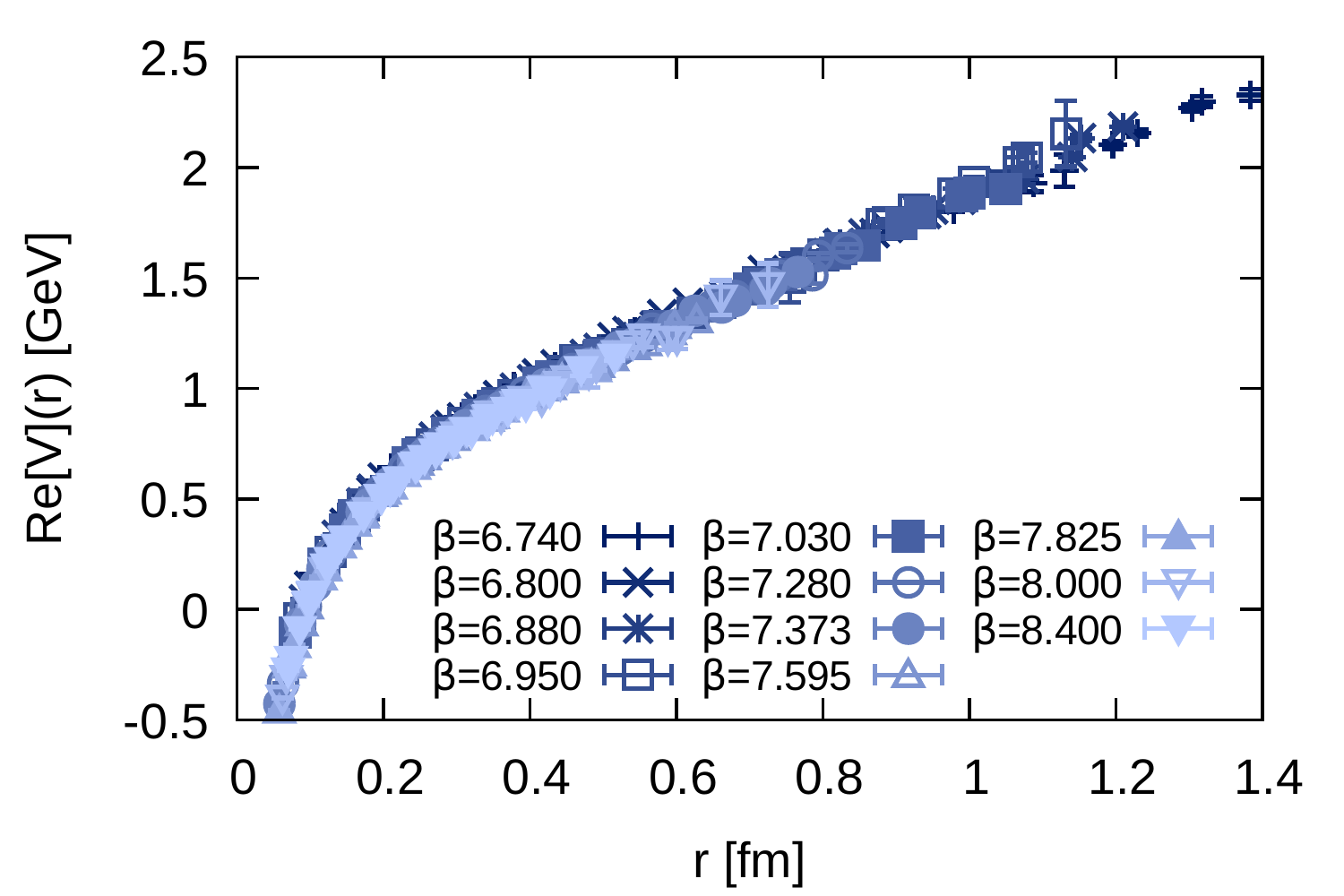}
  \end{center}\vspace{-0.7cm}
 \caption{The $T=0$ potential at different $\beta$'s.}\label{Fig:T0Pot}
\end{wrapfigure} 
\indent At $T=0$, we have $N_{\rm conf}=800-1200$ configurations per lattice spacing available with lattice extent between $N_\tau=32-64$. We deploy both the BR and the Pad\'e-like method to extract the spectra and find that both methods yield a well defined peak, whose position agrees with that found by a naive multi-exponential fit. A small artificial numerical width is observed in the reconstructions, which we take as null-baseline for the finite temperature reconstructions. $V_{\rm QCD}^{T=0}(r)$ is shown in Fig.\ref{Fig:T0Pot}, exhibiting the characteristic Cornell type form.

At finite temperature the reconstruction becomes more difficult, since now the lattices are only of $N_\tau=12$ and $N_\tau=16$ size and at the same time the accessible physical Euclidean time range is diminished. In addition, above $T_c$ the spectral peak defining the potential will broaden and it is known that the BR method will thus eventually display ringing artifacts that make a quantitative extraction of the peak position unreliable. \begin{wrapfigure}[30]{r}{0.45\textwidth}\vspace{-1.1cm}
  \begin{center}
     \includegraphics[scale=0.45]{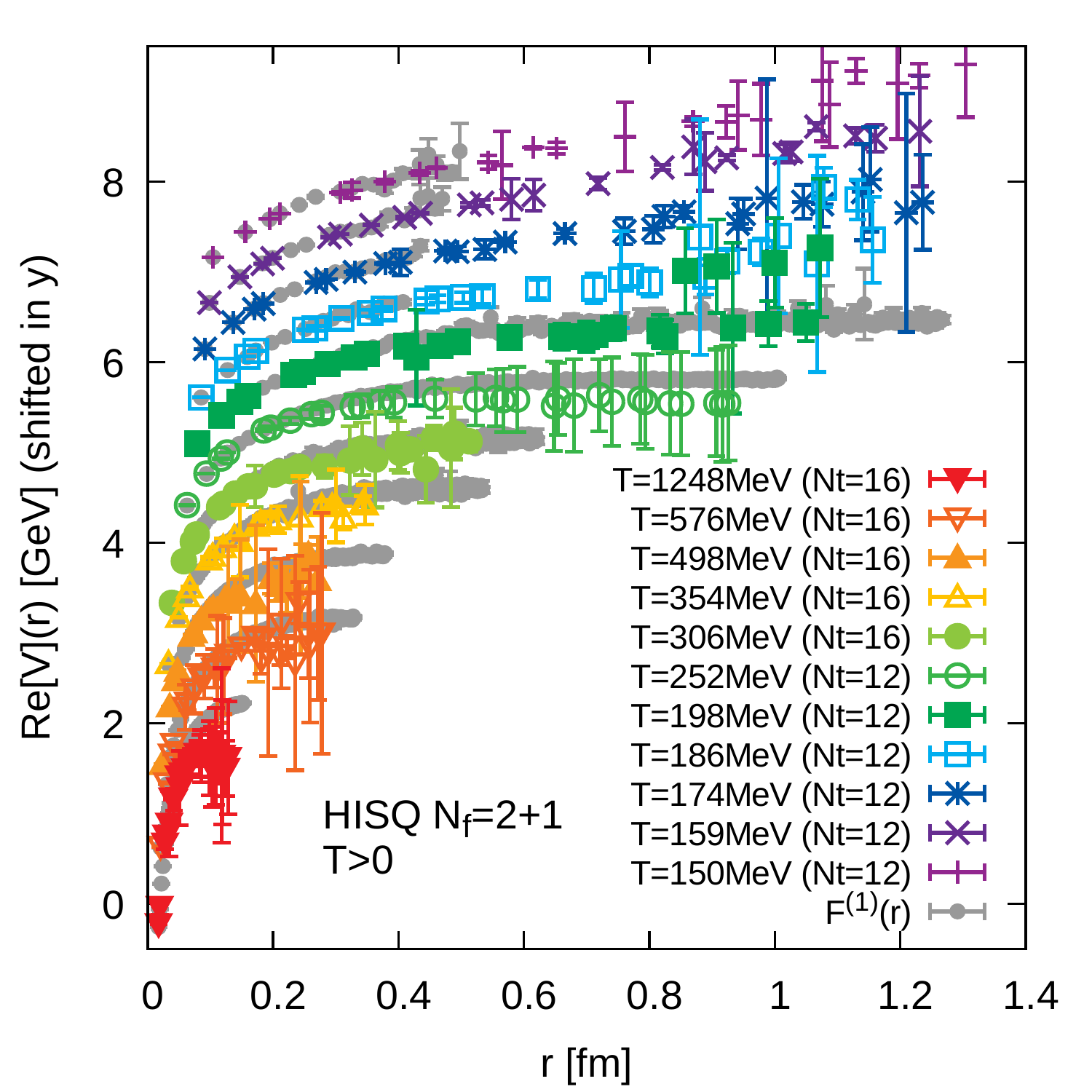}\vspace{-0.2cm}
     \includegraphics[scale=0.45]{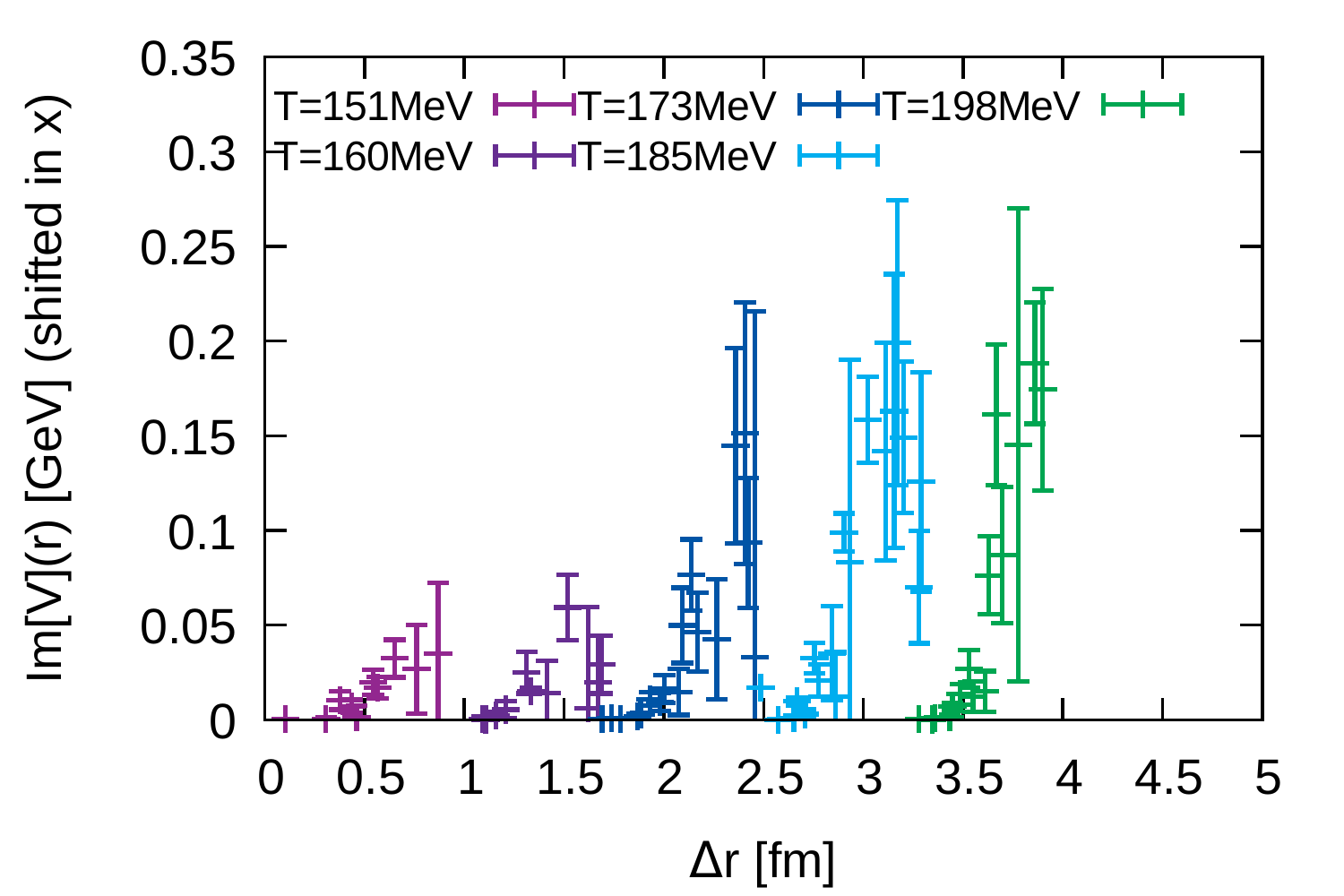}
  \end{center}\vspace{-0.7cm}
 \caption{(top) $\Re[V]$ via the Pad\'e-like method (colored), compared to the color-single free energies $F^1(r)$ in gray (shifted by hand in y-direction). (bottom) tentative values of $\Im[V]$ via the BR method (shifted by hand in x-direction).}\label{Fig:FiniteTPot}
\end{wrapfigure} 
Up to $T=198$MeV both BR and the Pad\'e-like method work reliably and provide the same results for $\Re[V]$. Starting with $T=252$MeV the BR method displays ringing and we only use the Pad\'e. In Fig.\ref{Fig:FiniteTPot} we thus show only the results from the Pad\'e extraction, where the uncertainty estimates arise from a 10-bin Jackknife and the variation among reconstructions where the input data is truncated at the latest available Euclidean times. The results on $\Re[V]$ in Fig.\ref{Fig:FiniteTPot} corroborate a qualitative picture consistent with previous studies \cite{Burnier:2014ssa} on less realistic lattices: The real part smoothly transitions from a Cornell-type form to a screened, i.e. asymptotically flat behavior in the QGP at high temperatures. At the same time we find that $\Re[V]$ is at all temperatures compatible with the color singlet free energies in Coulomb gauge within its uncertainties. In particular we do not find indications that $\Re[V]$ rises more steeply than $F^{(1)}(r)$. As was found in \cite{Bazavov:2018wmo} $F^{(1)}(r)$ deviates less than $2\%$ from its $T=0$ form up to $r<0.3/T$. Note that while agreement between $\Re[V]$ and $F^{(1)}(r)$ is only expected at $T=0$ and $T\gg T_c$, the extracted values even around $T_c$ are very similar.

$\Re[V]$ has been analyzed on some of these lattices using non-Bayesian approaches in the past. It has been modelled e.g. by assuming that the Wilson line spectrum also non-perturbatively follows the skewed Breit-Wigner plus shoulder form of HTL perturbation theory, with deviations encoded in a rescaling of frequencies \cite{Bazavov:2014kva}. On the other hand an extraction has been proposed using the first and second moments of the correlator \cite{Petreczky:2017aiz}, which in case of a Gaussian spectral function can be unambiguously related to the real- and imaginary part. The outcome of both studies was that $\Re[V]$ remains steeper than $F^{(1)}$ and lies quite close to the $T=0$ behavior. On the lattice the shoulder structure of the spectrum at higher frequencies apparently deviates significantly from the HTL form and thus the models may be driven artificially to higher values. While spectral reconstructions appear to more cleanly separate the shoulder from the actual potential peak contribution, further study is needed. 

The mock analysis showed that the Pad\'e-like method is unable to capture $\Im[V]$ in case of $N_\tau=12$ input datapoints. Here we instead perform an estimation via the BR method at $T\leq 198$MeV, where it is applicable. The lower panel in Fig.\ref{Fig:FiniteTPot} contains the tentative values, where the artificial numerical width present at $T=0$ has already been subtracted. Once temperatures rise above $T\approx 160$MeV we observe the presence of values in $\Im[V]$ that are significantly different from zero.

We are currently generating lattices with $N_\tau=16$ to reach similarly high statistics as at $N_\tau=12$. Subsequently an investigation of the screening of $\Re[V]$ using the Gauss-Law parametrization \cite{Burnier:2015nsa} will be performed. Efforts are underway to clarify how the EFT potential defined here is related to the quantity governing the dynamics in the T-matrix approach of \cite{Liu:2017qah}. We are confident that $V_{\rm QCD}^{T>0}(r)$, extracted non-perturbatively, will serve as vital input to phenomenological models (e.g. \cite{Brambilla:2017zei,Krouppa:2017jlg}), improving control over the evolution of heavy quarkonium in heavy-ion collisions from first principles. A.R. is supported by DFG via "SFB 1225 (ISOQUANT)", P.P. by DOE via DE-SC0012704 and J.H.W. by BMBF via 05P15WOCA1. We thank HotQCD and TUMQCD for their configurations.

\begin{multicols}{2}

\end{multicols}

\end{document}